\documentclass[10pt, twocolumn, twoside]{IEEEtran}
\usepackage{epsfig,latexsym}
\usepackage{float}
\usepackage{indentfirst}
\usepackage{amsmath}
\usepackage{amssymb}
\usepackage{xcolor}

\usepackage{times}
\usepackage{subfigure}
\usepackage{psfrag}
\usepackage{cite}
\usepackage{epstopdf}
\usepackage{fancyhdr}
\usepackage{stfloats}
\usepackage{color}
\newcommand{\OMA}{\mathrm{OMA}}

\newtheorem{theorem}{Theorem}

\begin{document}%
\title{On the Performance of NOMA with Hybrid ARQ}
 \vspace{-1.9em}
\author{Donghong~Cai, ~\IEEEmembership{Student Member,~IEEE,}~Zhiguo~Ding,~\IEEEmembership{Senior~Member,~IEEE,}\\
~Pingzhi Fan,~\IEEEmembership{Fellow,~IEEE,} and~Zheng Yang,~\IEEEmembership{Member,~IEEE}

\thanks{D. Cai and P. Fan are with the Institute of Mobile Communications, Southwest Jiaotong University,
Chengdu 610031, China (e-mail: cdhswjtu@163.com; pzfan@swjtu.edu.cn). Z. Ding is with the School of Electrical and Electronic Engineering, The University of Manchester, Manchester, UK, (e-mail: zhiguo.ding@manchester.ac.uk).  Z. Yang is with the Fujian Provincial Engineering Technology
Research Center of Photoelectric Sensing Application, Fujian Normal
University, Fuzhou 350007, China(email: zyfjnu@163.com).}
\thanks{The work of D. Cai
and P. Fan was supported by the National Science Foundation of China under Grant No. 61731017,  the National Science and Technology Major Project under Grant No. 2016ZX03001018-002 and the 111 Project under Grant No. 111-2-14. The work of Z. Ding was supported by the UK EPSRC under grant number EP/ P009719/1 and by H2020-MSCA-RISE-2015 under grant number 690750. The work of Z. Yang  was supported in part by the National Natural Science Foundation
of China under Grant 61701118 and Grant 61571128, and in part by
the Natural Science Foundation of Fujian Province, China, under Grant
2018J05101.}


	}

\maketitle
\begin{abstract}
In this paper, we investigate the outage performance of hybrid automatic repeat request with chase combining (HARQ-CC) assisted downlink non-orthogonal multiple access (NOMA) systems. A closed-form expression of the individual outage probability and the diversity gain  are obtained firstly. Based on the developed analytical outage probability, a tradeoff between the  minimum number of retransmissions and the transmit power allocation  coefficient is then provided for a given target rate. The provided simulation results demonstrate the accuracy of the developed analytical results. Moreover, it is shown that NOMA combined with the HARQ-CC can achieve a significant advantage when only average channel state information is known at the transmitter. Particularly, the performance of the user with less transmit power in NOMA systems can be efficiently improved by utilizing HARQ-CC.
\end{abstract}

\vspace{-0.5em}
\section{Introduction}
\vspace{0.5em}
\IEEEPARstart{C}ompared with conventional orthogonal multiple access (OMA), such as time division multiple access (TDMA) and frequency division multiple access (FDMA), non-orthogonal multiple access (NOMA) technique allows multiple users share the same resource block, i.e., time/frequency, by using different power levels and/or codes\cite{Dingmagazine2017, WenchenSCMA,CAIWCL}. Meanwhile, successive interference cancellation (SIC) is applied  to distinguish different users'  signals at the receiver side. Hence, NOMA achieves a significant performance improvement in terms of spectral efficiency and provides a large number of connections. Recently, the power allocation and the impact of channel state information (CSI) of NOMA have attracted
much attention \cite{Ding2014spl,DPATWC2016,SPL2015,Cuispl2016}. Specially,  in order to  guarantee  the  quality  of  service  (QoS) for  each  user in NOMA, a novel dynamic power allocation scheme has been proposed in \cite{DPATWC2016}, where the power allocation coefficient was determined by the instantaneous CSI. However, this assumption is impractical owing to the limited CSI feedback. Inspired by this, an optimal power allocation scheme was investigated to maximize the fairness among users has been investigated in \cite{SPL2015}, in which both instantaneous CSI and average CSI are considered. Furthermore, the authors in \cite{Cuispl2016} have optimized the transmit power and the decoding order of SIC with average CSI. These results reveal that the performance of NOMA will be degraded when only the average CSI is known by the transmitter.

In order to improve the performamce of NOMA with average CSI, hybrid automatic repeat request (HARQ) technique has been considered, as in \cite{HARQNOMAPIMRC,Choi2008,Choitcom2016}. For example,
in \cite{Choi2008}, the HARQ with chase combining
(HARQ-CC) protocol has been utilized to NOMA,  the obtained
results showed that the performance of NOMA outperforms that of OMA with flexible retransmission strategies. Moreover, the upper bound of outage probability and the power allocation of HARQ with incremental redundancy (HARQ-IR) aided NOMA have been considered in \cite{Choitcom2016}. However, there is still a lack of theoretic performance analysis of HARQ aided NOMA systems due to the  cumulative distribution function (CDF) of the sum of multiple signal-to-interference-plus-noise-ratios (SINRs) is difficult to evaluate.

In this paper, we consider the performance of a two-user downlink NOMA with the aid of HARQ-CC transmission protocol. Unlike the upper bound in \cite{Choitcom2016}, which requires the number of transmission rounds to be large enough and is not tight at all signal-to-noise-ratio (SNR) regime, we derive the individual outage probability of HARQ-CC assisted NOMA in closed-form expressions. It is proved that the analytical result matches well with the simulation result for all SNR regime and the performance of NOMA with HARQ-CC outperforms that of OMA with HARQ-CC, even the number of retransmission rounds is small. Furthermore, the diversity order is obtained at high SNRs, which is proportional to the number of transmission rounds. In addition,  the relationship between the minimum number of retransmissions and the power allocation coefficient for a given target rate is discussed, which provides insights: i)
for a given power allocation scheme, we can find the minimum number of
retransmissions to guarantee the QoS requirement. ii) with a fixed
number of retransmissions, the minimum transmit power coefficient can be
found for satisfying the QoS requirement.

\vspace{-0.5em}
\section{System Model}
\vspace{0.5em}
Consider a two-user downlink NOMA system\footnote{A two-user downlink NOMA,
named as multi-user superposition transmission (MUST), has
been studied for the Third Generation Partnership Project
(3GPP) Long Term Evolution (LTE)\cite{li2013energy}.}, where the users and the base station (BS) are modelled with a single-antenna. The observation at the $j$-th user during the $t$-th transmission round can be represented as follows:
\begin{align}\label{eqn:mod1}
y_{jt}=h_{jt}(\sqrt{\alpha_1P}s_{1t}+\sqrt{\alpha_2P}s_{2t})+n_{jt},
\end{align}
where $j\in\{1,2\}, s_{jt}$ is the signal of user $j$ with unit-energy; $h_{jt}=\frac{g_{jt}}{\sqrt{1+d_j^{\zeta}}}$ denotes the channel coefficient between the BS and user $j$ during the $t$-th transmission round, $\zeta$ denotes the path loss exponent, $d_j$ is the distance, and $g_{jt}\thicksim \mathcal{CN}(0,1)$; the total transmit power is $P$; $n_{jt}$ is additive white Gaussian noise (AWGN) with variance $\sigma^2$. The power allocation factors for the two users are $\alpha_1$ and $\alpha_2$, respectively, where $\alpha_j\in (0,1), \alpha_1+\alpha_2=1$. It is assumed that $E[|h_{1t}|^2]\le E[|h_{2t}|^2]$, i.e., $d_1\geq d_2$. According to NOMA principles,  user 1 should be allocated more power for the fairness, i.e., $\alpha_1>\alpha_2$.
Thus, SIC is always applied at user 2 to decode the signal of user 1 firstly, where the received SINR of user 2 to decode the signal of user 1 and the received SNR to decode its own signal can be expressed as
\begin{align}\label{SINRSNR}
\gamma^t_{2\rightarrow1}=\frac{\alpha_1|h_{2t}|^2}{\alpha_2|h_{2t}|^2+1/\rho}, \gamma^t_{2\rightarrow2}=\alpha_2\rho |h_{2t}|^2,
\end{align}
respectively, where $\rho=\frac{P}{\sigma^2}$ denotes the transmit SNR.

At user 1, the signals of user 2 are always treated as interference, then the received SINR for decoding its own signal can be given by
\begin{align}\label{SINR}
    \gamma^t_{1\rightarrow1}=\frac{\alpha_1|h_{1t}|^2}{\alpha_2|h_{1t}|^2+1/\rho}.
\end{align}

In order to enhance the reliability of transmission in NOMA systems, the HARQ-CC protocol is employed. With the aid of HARQ-CC, each user in NOMA will store the signal and send a negative acknowledgment (NACK) signal to the BS whenever it fails to decode its own signal. Otherwise, an acknowledgment (ACK) will be fed back to the BS for requesting a new signal transmission. Only two ACK signals of two users are received at the BS at the same time, a new superimposed code will be sent to the users. If not,  the same superimposed code will be send to the users for decoding again. For the case that each user has a delay requirement, i.e. a maximum number of retransmission constraint, if someone still can not decode the signal, the superimposed code will be discarded. Moreover, maximum ratio combining (MRC) is used at the receivers, which combines the signal of retransmission with the previously failed signal for decoding. In this way, after $T$ transmission rounds, the outage probabilities of user 1 and user 2 in NOMA can be formulated as
\begin{align}\label{eqn:mod6}
P^N_{1,T}=\Pr\bigg\{\log_2\bigg(1+\sum_{t=1}^T \gamma^t_{1\rightarrow1}\bigg)\leq T\widehat{R}_1 \bigg\},
\end{align}
and
\begin{align}\label{eqn:mod7}
P^N_{2,T}=&1-\Pr\bigg\{\log_2\bigg(1+\sum_{t=1}^T \gamma^t_{2\rightarrow1}\bigg)> T\widehat{R}_1,\nonumber\\
&\log_2\bigg(1+\sum_{t=1}^T \gamma^t_{2\rightarrow2}\bigg)> T\widehat{R}_2\bigg\},
\end{align}
respectively, where $\widehat{R}_j$ is the target rate of the $j$-th user. For comparison, the conventional  orthogonal  multiple  access, such as TDMA, the outage probability of OMA with HARQ-CC can be expressed as
\begin{align}\label{eqn:mod8}
P_{j,T}^{\OMA}=\Pr\bigg\{\frac{1}{2}\log_2\bigg(1+\sum_{t=1}^T\rho|h_{jt}|^2 \bigg)\le T\widehat{R}_j\bigg\}.
\end{align}
\vspace{-0.5em}
\section{Performance Analysis}
\vspace{0.5em}
In this section, we first focus on  the individual outage probability
 and the diversity order of NOMA with the aid of HARQ-CC transmission protocol. Furthermore, based on the developed outage probability, we study the relationship between the number of retransmissions and power allocation coefficient with QoS requirement constraint.
\subsection{Individual Outage probability}
Recall that the CDF of $\gamma^t_{j\rightarrow 1}$ can be given by
\begin{align}\label{eqn:AD1}
F_{\gamma^t_{j\rightarrow 1}}(y_t)&=\Pr\bigg(\frac{\alpha_1|h_{jt}|^2}{\alpha_2|h_{jt}|^2+1/\rho}< y_t\bigg)\nonumber\\
&=\Pr\bigg((\alpha_1-\alpha_2y_t)|h_{jt}|^2< \frac{y_t}{\rho}\bigg).
\end{align}
Since $|h_{jt}|^2$ follows an exponential distribution, we have
\begin{align}\label{eqn:cdfsnin1}
    F_{\gamma^t_{j\rightarrow 1}}(y_t)=1-e^{-\frac{y_t}{\lambda_1\rho(\alpha_1-\alpha_2y_t)}},
\end{align}
for $0<y_t<\frac{\alpha_1}{\alpha_2}$, where $\lambda_1=\frac{1}{1+d_1^{\zeta}}$.\\
And
\begin{align}
F_{\gamma^t_{j\rightarrow 1}}(y_t)=1,
\end{align}
for $y_t\geq\frac{\alpha_1}{\alpha_2}$.
Note that the outage probability of user $j$ can be expressed as $P_{j,1}^N= F_{\gamma^T_{j\rightarrow 1}}(r_j)$ for $T=1$, where $r_j=2^{\widehat{R}_j}-1$. And $P_{j,1}^N=1$ for $r_1\geq\frac{\alpha_1}{\alpha_2}$. Hence, we only consider the individual outage probability of HARQ-CC aided NOMA with $0<r_1<\frac{\alpha_1}{\alpha_2}$ in the following theorems.
\begin{theorem} \label{thm:outage}
The outage probability of user 1 in NOMA with HARQ-CC systems can be approximated as follows:
\begin{align}\label{eqn:PA12}
P_{1,T}^{N}&\approx c^T\sum_{\{j_1,...,j_N\}\in\Phi}\eta\bigg[\prod_{n=1}^N\mathcal{C}^{j_n}(x_n)\bigg] \sum_{m=1}^N \frac{\pi\ln2\sqrt{1-x_m^2}}{Nr_1(1+x_{m})}\nonumber\\ &\cdot\left[\sum_{k=1}^L\hat{w}_ke^{-\frac{\sum_{n=1}^Nj_n\beta(x_n+1)k\ln2}{r_1(1+x_m)}}\right],
\end{align}
where $x_i=\cos\bigg(\frac{2i-1}{2N}\pi\bigg), i=n,m$, $\eta=\frac{T!}{\prod_{n=1}^Nj_n!},$ $\beta=\frac{\alpha_1}{\alpha_2},$ $c=\frac{2\alpha_1\beta\pi}{N\lambda_1\rho}e^{\frac{1}{\alpha_2\lambda_1\rho}}$, $ 0<r_1<\beta$, the set $\Phi$ is defined as follows:
$$
\Phi=\bigg\{j_1,...,j_N|T=\sum_{n=1}^Nj_n, j_n=0,1,...,T\bigg\},
$$
the functions, $\mathcal{C}(x), \hat{w}_k$, are defined as
\begin{align}\label{ccfunction}
\mathcal{C}(x)=\frac{\sqrt{1-x^2}}{(2\alpha_1-\alpha_2\beta(x+1))^2}e^{-\frac{2\alpha_1}{\alpha_2\lambda_1\rho(2\alpha_1-\alpha_2\beta(x+1))}},
\end{align}
and
\begin{align*}
    \hat{w}_k=(-1)^{\frac{L}{2}+k}\sum_{l=\left \lfloor \frac{k+1}{2} \right \rfloor}^{\min\{k,l\}}\frac{l^{L/2+1}}{(L/2)!}\binom{L/2}{l}\binom{2l}{l}\binom{l}{k-l},
\end{align*}
respectively, $\left \lfloor k \right \rfloor$ denotes the largest integer not exceeding $k$; $N$ and $L$ are parameters to ensure a complexity-accuracy tradeoff.
\end{theorem}

From the closed-form expression of outage probability for user 1 in Theorem 1, we can see that each term of \eqref{eqn:PA12} tends to zero when the SNR is large enough. Thus, the signal of user 1  at user 2 can also be decoded with a high success probability, and the outage probability of user 2 can be approximated in the following theorem.

\begin{theorem} \label{thm:outageu2}
In NOMA with HARQ-CC, the outage probability of user 2 at high SNR regime can be approximated as the follows
\begin{align}\label{eqn:PA22}
P_{2,T}^{N}\approx \gamma\bigg(T,\frac{2^{T\widehat{R}_2}-1}{\alpha_2\lambda_2\rho}\bigg),
\end{align}
with
$$
\frac{1}{T}\log_2\bigg(\frac{\alpha_1T}{\alpha_2}+1\bigg)>\widehat{R}_1.
$$
where $\lambda_2=\frac{1}{1+d_2^{\zeta}}$ and $\gamma(\alpha,x)=\int_0^x e^{-s}s^{\alpha-1}ds$ is the incomplete gamma function \cite{GRADSHTEYN}.
\end{theorem}
\begin{IEEEproof}
Since the SINR in \eqref{SINRSNR} can be expressed as
$$
\sum_{t=1}^T\frac{\alpha_1\rho|h_{2t}|^2}{\alpha_2\rho|h_{2t}|^2+1}=\frac{\alpha_1}{\alpha_2}\sum_{t=1}^T\bigg(1-\frac{1}{\alpha_2\rho|h_{2t}|^2+1}\bigg),
$$
then the outage probability of user 2 in \eqref{eqn:mod7} can be equivalently reformulated as
\begin{align}\label{eqn:PA3}
P^N_{2,T}=&1-\Pr\bigg\{\sum_{t=1}^T\frac{1}{\alpha_2\rho|h_{2t}|^2+1}<T- \frac{\alpha_2(2^{T\widehat{R}_1}-1)}{\alpha_1},\nonumber\\
&\sum_{t=1}^T(\alpha_2\rho|h_{2t}|^2+1)> 2^{T\widehat{R}_2}+T-1\bigg\}.
\end{align}
Note that $\frac{1}{\alpha_2\rho|h_{2t}|^2+1}\leq\alpha_2\rho|h_{2t}|^2+1$, and $\frac{1}{\alpha_2\rho|h_{2t}|^2+1}\rightarrow 0$, $\alpha_2\rho|h_{2t}|^2+1\rightarrow\infty$ for $\rho\rightarrow\infty$. It means that the signals of user 1 can be decoded and removed perfectly at user 2 at high SNR regime.

Thus, when the target rate of user 1 meets the following condition
$$
\frac{1}{T}\log_2\bigg(\frac{\alpha_1T}{\alpha_2}+1\bigg)>\widehat{R}_1.
$$
The outage probability of user 2 in \eqref{eqn:PA3} can be approximated as
\begin{align}\label{eqn:PA4}
P^N_{2,T}&\approx\Pr\bigg\{\log_2\bigg(1+\sum_{t=1}^T \gamma^t_{2\rightarrow2}\bigg)\leq T\widehat{R}_2\bigg\}\nonumber\\
&=\Pr\bigg\{\sum_{t=1}^T \gamma^t_{2\rightarrow2}\leq 2^{T\widehat{R}_2}-1\bigg\}\nonumber\\
&\overset{(a)}{=}\gamma\bigg(T,\frac{2^{T\widehat{R}_2}-1}{\alpha_2\lambda_2\rho}\bigg),
\end{align}
where step $(a)$ follows the fact that the sum of $T$ independentand identically distributed (i.i.d.) SINRs follows a center chi-square distribution with degrees of freedom $T$.
\end{IEEEproof}

\subsection{Diversity Order}
With the aid of the developed outage probability in Theorem 1 and Theorem 2, the diversity order of NOMA with HARQ-CC can be derived in this subsection.

By defining
$$
\mu= \sum_{m=1}^N \frac{\pi\ln2\sqrt{1-x_m^2}}{Nr_1(1+x_{m})}\left[\sum_{k=1}^L\hat{w}_ke^{-\frac{\sum_{n=1}^Nj_n\beta(x_n+1)k\ln2}{r_1(1+x_m)}}\right],
$$
the diversity order of user 1 in NOMA with HARQ-CC can be given by
\begin{align}\label{eqn:PA121}
\mu_1&=\lim_{\rho \to \infty}\frac{\log\bigg(c^T\sum_{\{j_1,...,j_N\}\in\Phi}\eta\bigg[\prod_{n=1}^N\mathcal{C}^{j_n}(x_n)\bigg]\mu\bigg)}{-\log(\rho)}\nonumber\\
&=-\underbrace{\lim_{\rho \to \infty}\frac{\log\bigg(\sum_{\{j_1,...,j_N\}\in\Phi}\eta\bigg[\prod_{n=1}^N\mathcal{C}^{j_n}(x_n)\bigg]\mu\bigg)}{\log(\rho)}}_{Q_1}\nonumber\\
&-\lim_{\rho \to \infty}\frac{\log(c^T)}{\log(\rho)}.
\end{align}
From \eqref{ccfunction}, we have $$\lim_{\rho \to \infty}\mathcal{C}(x)=\frac{\sqrt{1-x^2}}{(2\alpha_1-\alpha_2\beta_1(x+1))^2},$$
then $Q_1=0$ in \eqref{eqn:PA121}.  Further, the diversity order of user 1 can be  simplified to
\begin{align}
\mu_1&=-\lim_{\rho \to \infty}\frac{\ln(\frac{1}{\rho^T})+T\ln(\frac{2\alpha_1\beta_1\pi}{N\lambda_1})+\frac{T}{\alpha_1\lambda_2\rho}}{\ln(\rho)}=T.
\end{align}
For user 2, the incomplete gamma function, $\gamma(\alpha,x)$, in \eqref{eqn:PA22} can be expressed as the following series representation
\begin{align}\label{imcom}
\gamma(\alpha,x)=\sum_{n=0}^{\infty}\frac{(-1)^{n}x^{\alpha+n}}{n!(\alpha+n)}.
\end{align}
From \eqref{eqn:PA22} and \eqref{imcom}, the diversity order of user 2 can be given by
\begin{align}
\mu_2&=-\lim_{\rho \to \infty}\frac{\ln(\frac{1}{\rho^T})+T\ln(\frac{2^{T\widehat{R}_2}-1}{\alpha_2\lambda_2})+\ln(\sum_{n=0}^{\infty}\frac{(-x_0)^{n}}{n!(T+n)})}{\ln(\rho)}\nonumber\\
&=T,
\end{align}
where $x_0=\frac{2^{T\widehat{R}_2}-1}{\alpha_2\lambda_2\rho}$. It is important to point out that the diversity orders of user 1 and user 2 are proportional to the number of transmission rounds. In order to further reveal the impact of the retransmission, we will discuss the relationship of retransmission rounds and the power allocation in the next subsection.

\subsection{A tradeoff between retransmission times and power allocation coefficient}

Similar to the analysis of user 2 in NOMA with HARQ-CC, the
individual outage probability of OMA with HARQ-CC in \eqref{eqn:mod8} can
be easily obtained as
\begin{align}\label{eqn:PA521}
P_{j,T}^{\OMA}=\gamma\bigg(T,\frac{2^{2T\widehat{R}_j}-1}{\lambda_j\rho}\bigg).
\end{align}
 Moreover, the diversity order of each user in OMA with HARQ-CC
can be easily obtained based on the derived outage probability, which is also equal to
$T$.

Recall that the outage probabilities of NOMA and OMA for each transmission round can be given by
\begin{align}\label{nomaout1}
    P_{1,1}^{N}=1-e^{-\frac{2^{\widehat{R}_1}-1}{(\alpha_1-\alpha_2(2^{\widehat{R}_1}-1))\rho}},
\end{align}
and
\begin{align}\label{omaout111}
    P_{1,1}^{\OMA}=1-e^{-\frac{2^{2\widehat{R}_1}-1}{\rho}},
\end{align}
respectively. Based on the outage probabilities in \eqref{nomaout1} and \eqref{omaout111}, we can see that $P_{1,1}^{N}<P_{1,1}^{\OMA}$ if and only if $\frac{\alpha_1}{\alpha_2}>2^{\widehat{R}_1}> 1$, which  always holds for the case that the fairness among users is considered in NOMA. Therefore, the outage performance of user 1 in NOMA with HARQ-CC outperforms that of OMA with HARQ-CC when more power is allocated to user 1. However, in this case, user 2 with less transmit power might  not  be  able  to  meet  the QoS requirement.

From \eqref{eqn:PA4} and \eqref{eqn:PA521}, in order to guarantee the QoS of user 2 in NOMA with HARQ-CC, we should have
\begin{align}
\gamma\bigg(T,\frac{2^{T\widehat{R}_2}-1}{\alpha_2\lambda_2\rho}\bigg)<\gamma\bigg(T,\frac{2^{2T\widehat{R}_2}-1}{\lambda_2\rho}\bigg).
\end{align}
Then a tradeoff between $\alpha_2$ and $T$ with given $\widehat{R}_2$ can be discussed as the following two cases:

i) Given $\alpha_2$ and $\widehat{R}_2$, we can define a function on $T$ as follows:
\begin{align}\label{eqn:PA6}
K(T|\alpha_2,\widehat{R}_2)=\frac{2^{T\widehat{R}_2}-1}{\alpha_2}-2^{2T\widehat{R}_2}+1,
\end{align}
in which $K(T|\alpha_2,\widehat{R}_2)< 0$ denotes that NOMA with HARQ-CC outperforms that of OMA with HARQ-CC.
Then the minimum retransmission round of HARQ-CC can be given by
\begin{align}\label{eqn:PA712}
 \widehat{T}=\min\ \left\{T \bigg| \frac{\log_2(\frac{1-\alpha_2}{\alpha_2})}{\widehat{R}_2}<T\right\}.
\end{align}

ii) Given $\widehat{R}_2$ and $T$, in order to guarantee the QoS of user 2, we should find an $\alpha_2$ such that \begin{align}\label{eqn:PA621}
L(\alpha_2|\widehat{R}_2, T)\triangleq\frac{2^{T\widehat{R}_2}-1}{\alpha_2}-2^{2T\widehat{R}_2}+1<0.
\end{align}
From \eqref{eqn:PA621}, the transmit power allocated to user 2 should meet the following condition
\begin{align}\label{powercondition}
\alpha_2>\sup \frac{2^{T\widehat{R}_2}-1}{2^{2T\widehat{R}_2}-1},\ \text{for}\  \widehat{R}_2\in(0, +\infty),
\end{align}
where $\sup$ denotes the supremum.

For a special case, $T=1$, the power coefficient of user 2 is $\alpha_2\in(\sup G(\widehat{R}_2),1)$, where $G(\widehat{R}_2)=\frac{2^{\widehat{R}_2}-1}{2^{2\widehat{R}_2}-1}$.

Since $\frac{\partial G(\widehat{R}_2)}{\partial \widehat{R}_2}<0,$
we have
$$\sup G(\widehat{R}_2)=\lim_{\widehat{R}_2\rightarrow 0}G(\widehat{R}_2)=0.5.$$

It means that the performance of user 2 in NOMA will better than that of OMA when the transmit power of this user is larger than 0.5. However, this condition contradicts the fairness of NOMA.

\vspace{-0.5em}
\section{Numerical Results And Simulations}
\vspace{0.5em}
In order to  verify the accuracy of the developed analytical results and compare the performance between NOMA and OMA, Monte Carlo simulation results are provided in this section. In all simulations, we set $N=20, L=10$ and the distances between the users and the BS are $d_1=7m$ and $d_2=3m$, respectively.

\begin{figure}[tp]
 \begin{center}
 \vspace*{-0.5em}{
\includegraphics[width=1.0\linewidth]{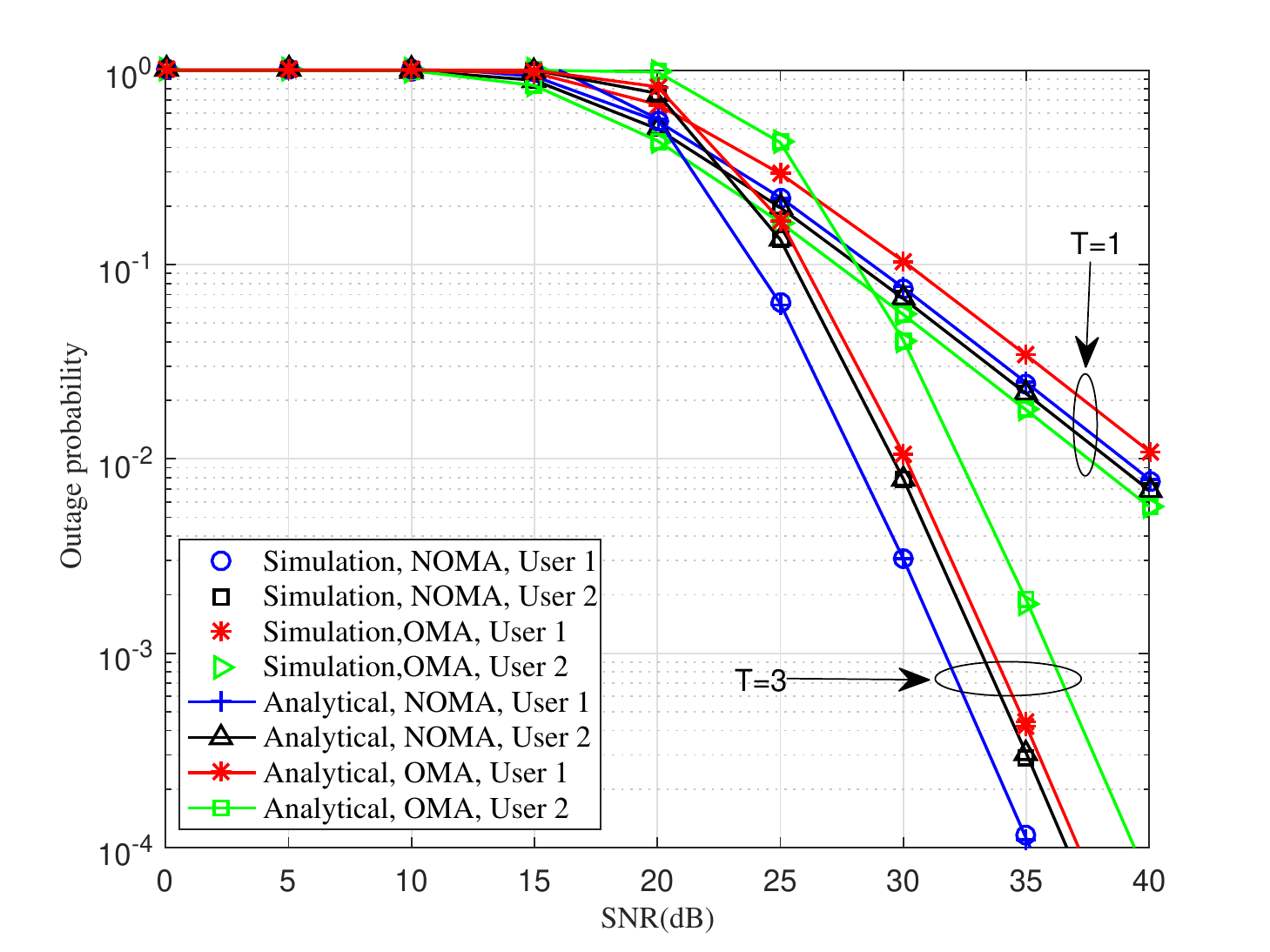}}
\end{center}
\vspace*{-1.5em}
\caption{ The outage probability comparison of NOMA with different $T$, where $\alpha_{1}=0.7, \alpha_{2}=0.3, \zeta=3, \widehat{R}_1=0.2~ \text{bps/Hz}, \widehat{R}_2=0.8~ \text{bps/Hz}$}.\label{FIG1}\vspace{-0.5em}
\end{figure}
\begin{figure}[tp]
 \begin{center}
 \vspace*{-0.5em}{
\includegraphics[width=1.0\linewidth]{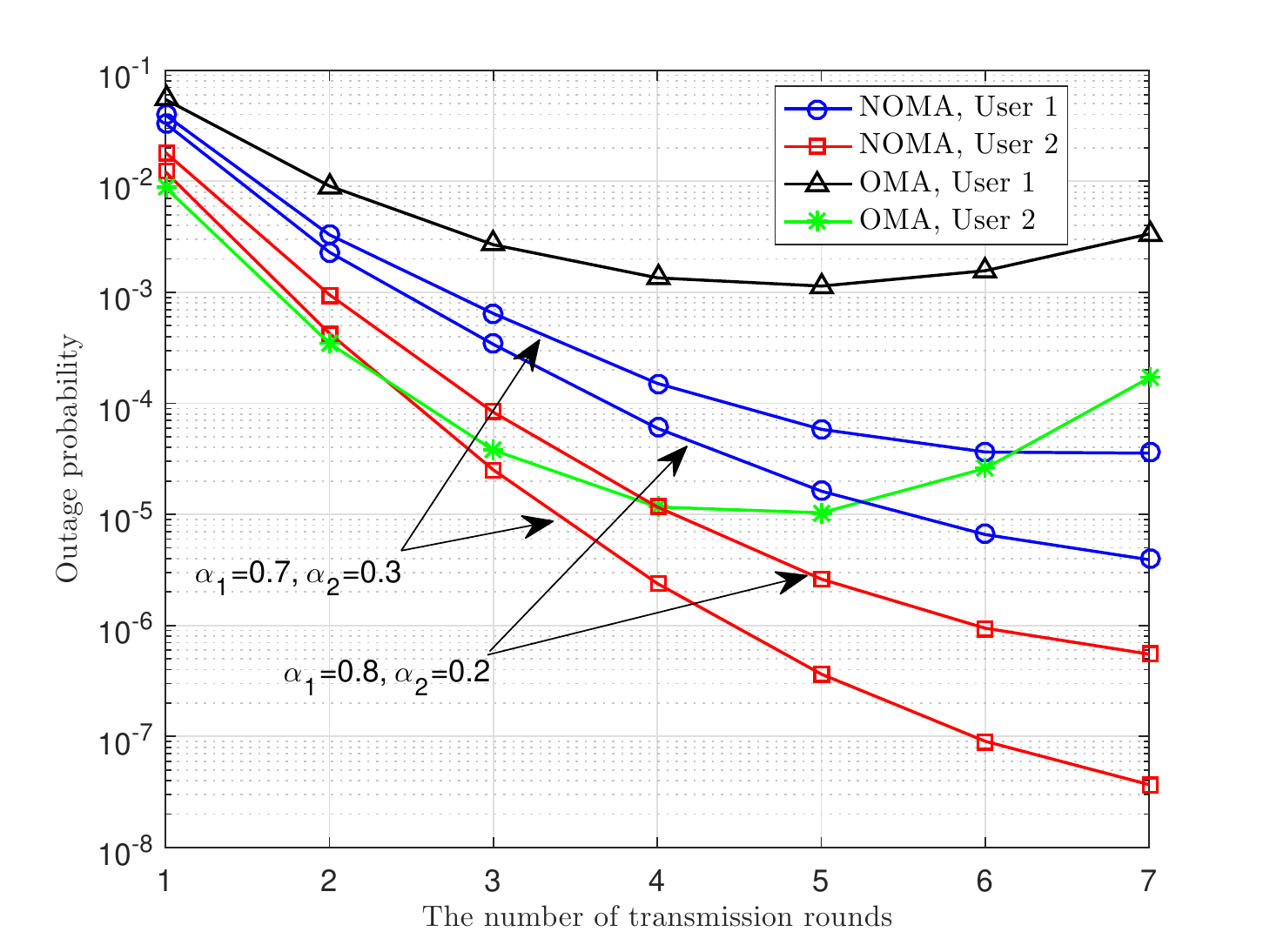}}
\end{center}
\vspace*{-1.5em}
\caption{The outage probability versus the number of transmission rounds, where $\zeta=3, \widehat{R}_1=0.3 ~\text{bps/Hz}, \widehat{R}_2=0.5~ \text{bps/Hz}, \mathrm{SNR}=35 \mathrm{dB}$}.\label{FIG4}\vspace{-0.5em}
\end{figure}
Fig. 1 plots the outage probabilities  of NOMA and OMA with different transmission rounds, in which $T=1$ corresponds to the case that NOMA without HARQ-CC transmission protocols. One can observe that the analytical results match the simulation results well. Also, we can find that the performance of user 1 in NOMA always outperforms that of OMA. However, user 2 in OMA achieves a better performance than NOMA for $T=1$. This is because that the power of user 2 can not meet the condition in \eqref{powercondition} for a given target rate. Moreover, it can be seen that the performance of NOMA can be improved by using HARQ-CC with flexible  retransmission  rounds.

In Fig. 2, the number of transmission rounds versus the outage probability is shown, where two power allocation schemes are considered, i.e., $\alpha_1=0.7, \alpha_2=0.3$ and $\alpha_1=0.8, \alpha_2=0.2$. We can see that the outage performance of user 1 in NOMA with HARQ-CC is better than that of OMA with HARQ-CC for all transmission rounds. However, for $T=1$, OMA achieves a better performance than NOMA. The reason is that power allocation schemes can only guarantee the QoS requirement of user 1 with the fixed target rate.   Besides, the outage performance of NOMA can be improved when we increase the number of transmission rounds. Especially, the minimum transmission rounds of these two power allocation schemes are 3 and 4 according to the  result in \eqref{eqn:PA712}, which are also verified in Fig. 2.

\begin{figure}[tp]
 \begin{center}
 \vspace*{-0.5em}{
\includegraphics[width=0.9\linewidth]{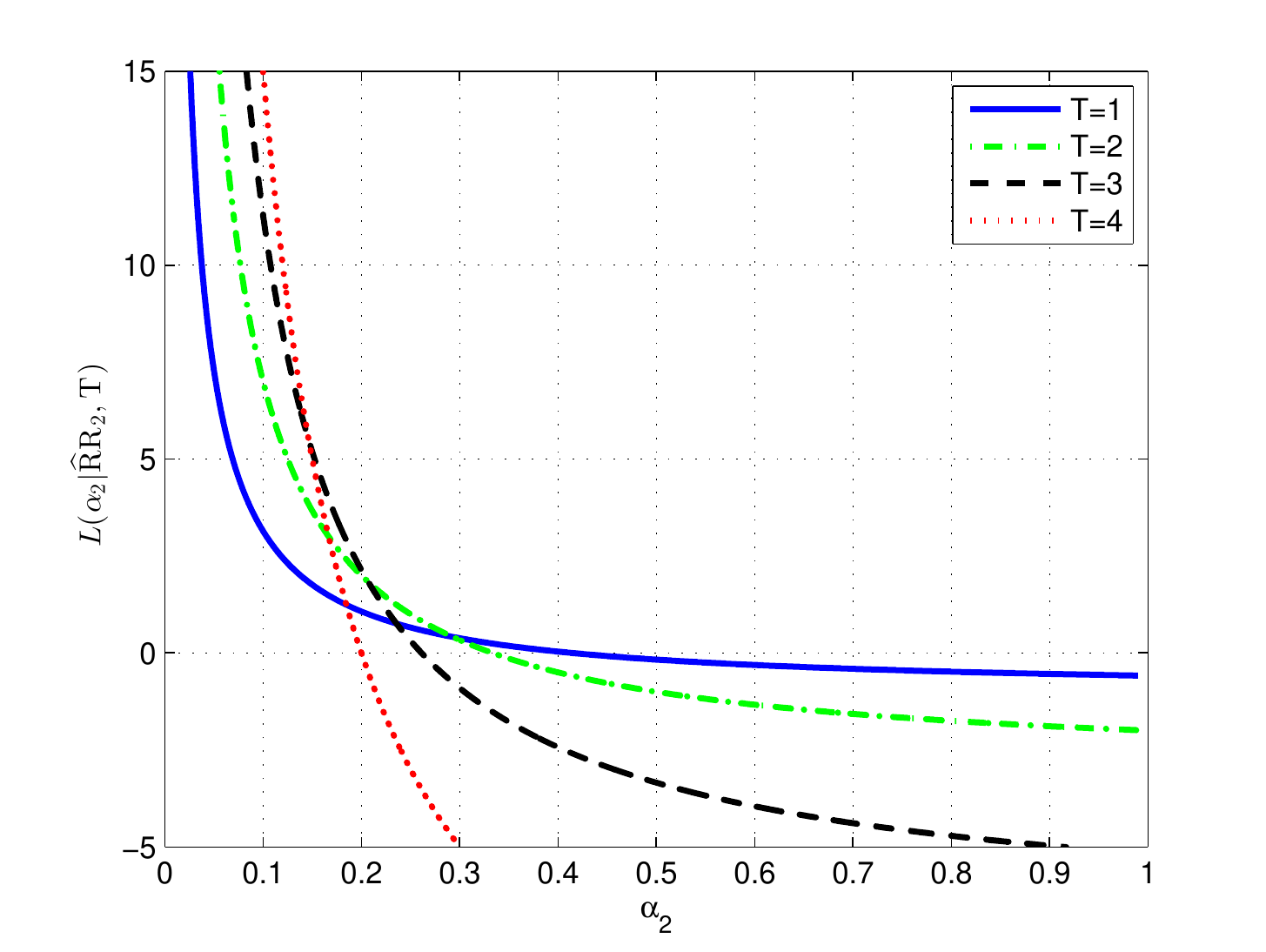}}
\end{center}
\vspace*{-1.5em}
\caption{$\alpha_2$ vs. $L(\alpha_2| \widehat{R}_2, T)$ of HARQ-CC aided NOMA with different $T$, where $\widehat{R}_2=0.5~ \text{bps/Hz}$.}\label{FIG421}\vspace{-0.5em}
\end{figure}
\begin{table}[tp]
\begin{center}
\caption{The values of $\alpha_2$ when $L(\alpha_2| \widehat{R}_2, T)=0$}  \label{tab1}
\begin{tabular}{|p{1cm}|p{1cm}|p{1cm}|p{1cm}|p{1cm}|}
  \hline
  T &1 &2&3&4 \\
   \hline
  $\alpha_2$& 0.4142 &0.3333&0.2612&0.2000\\
   \hline
\end{tabular}
\end{center}
\end{table}
In Fig. 3, the values of $L(\alpha_2| \widehat{R}_2, T)$ are shown as a function of $\alpha_2$ with given transmission round and target rate, where $L(\alpha_2| \widehat{R}_2, T)<0$   denotes that the outage performance of  NOMA outperforms that of OMA. Meanwhile, the minimum powers of user 2 are given in Table I for different transmission rounds, which guarantees the QoS requirement of this user. We can find that the power of user 2 decreases when the number of transmission rounds becomes large.

\vspace{-0.5em}
\section{Conclusions}
\vspace{0.5em}
In this paper, the  closed-form expression of individual outage probability for downlink NOMA combined with HARQ-CC transmission protocol has been derived. Furthermore, a tradeoff between the minimum transmission round and the power allocation scheme is discussed based on developed outage probability.  The developed analytical results and simulations have shown the superior outage performance of NOMA in the present of HARQ-CC, which can be used to guide the design of systems with delay and reliability requirements.

\appendices
\section{The Proof of Theorem 1}
\vspace{0.5em}

From \eqref{SINR}, the SINR of user 1 in NOMA with HARQ-CC after $T$ rounds, can be expressed as
\begin{align}\label{sumSINR1}
    Y=\sum_{t=1}^Ty_t,
\end{align}
where $\gamma^t_{1\rightarrow 1}\triangleq y_t$. Because the SINRs of different transmission rounds  are independent,
by defining $f_Y(y)\triangleq f(y_1,y_2,...,y_T)$, the joint PDF of $Y$ can be given by
\begin{align}\label{jointpdf}
f_Y(y)=\prod_{t=1}^Tf_{y_t}(y_t).
\end{align}

Applying Laplace transform to \eqref{jointpdf}, we have
\begin{align}\label{eqn:AD3}
&\mathcal{L}_Y(s)=\int_0^{+\infty}f_Y(y)e^{-sy}dy\nonumber\\
&=\int_0^{\infty}\cdots\int_0^{\infty}f(y_1,...,y_T)e^{-s(y_1+...+y_T)}dy_1...dy_T\nonumber\\
&=\int_0^{\infty}f(y_1)e^{-sy_1}dy_1\cdots\int_0^{\infty}f(y_T)e^{-sy_T}dy_T.
\end{align}
From \eqref{eqn:AD1} and \eqref{eqn:cdfsnin1}, the probability distribution function (PDF) of $y_t$ can be expressed as
\begin{align}\label{eqn:AD2}
f_{\gamma^t_{1\rightarrow 1}}(y_t)=\frac{\alpha_1}{\lambda_1\rho(\alpha_1-\alpha_2y_t)^2}e^{-\frac{y_t}{\lambda_1\rho(\alpha_1-\alpha_2y_t)}},
\end{align}
where $0<y_t<\frac{\alpha_1}{\alpha_2}$. Moreover, $\{y_t\}_{t=1}^T$ are
i.i.d. random variables.

Then, the Laplace transform of $Y$ in \eqref{eqn:AD3} can be equivalently reformulated
as
\begin{align}
    \mathcal{L}_Y(s)=\prod_{t=1}^T\mathcal{L}_{y_t}(s),
\end{align}
where $\mathcal{L}_{y_t}(s)$ denotes the Laplace transform of $y_t$ and is given by
\begin{align}\label{eqn:AD4}
\mathcal{L}_{y_t}(s)&=\int_0^{\beta}f(y_t)e^{-sy_t}dy_t.
\end{align}

By applying Gaussian-Chebyshev quadrature, \eqref{eqn:AD4} can be simplified as follows:
\begin{align}\label{eqn:AD5}
\mathcal{L}_{y_t}(s)&\approx c\sum_{n=1}^N\mathcal{C}(x_n)e^{-\frac{s\beta(x_n+1)}{2}},
\end{align}
where $x_n=\cos\bigg(\frac{2n-1}{2N}\pi\bigg)$, $c=\frac{2\alpha_1\beta\pi}{N\lambda_1\rho}e^{\frac{1}{\alpha_2\lambda_1\rho}}$, the function
$$\mathcal{C}(x)=\frac{\sqrt{1-x^2}}{(2\alpha_1-\alpha_2\beta(x+1))^2}e^{-\frac{2\alpha_1}{\alpha_2\lambda_1\rho(2\alpha_1-\alpha_2\beta(x+1))}},$$
and $N$ is a parameter to ensure a complexity-accuracy tradeoff.

Submitting \eqref{eqn:AD5} into \eqref{eqn:AD3}, the Laplace transform of $Y$ can be approximated as follows:
\begin{align}\label{eqn:AD6}
&[\mathcal{L}_{y_t}(s)]^T\approx c^T\bigg[\sum_{n=1}^N\mathcal{C}(x_n)e^{-\frac{s\beta(x_n+1)}{2}}\bigg]^T\nonumber\\
&=c^T\sum_{\{j_1,...,j_N\}\in\Phi}\eta\prod_{n=1}^N\mathcal{C}^{j_n}(x_n)e^{-\frac{sj_n\beta(x_n+1)}{2}}\nonumber\\
&=c^T\sum_{\{j_1,...,j_N\}\in\Phi}\eta\bigg[\prod_{n=1}^N\mathcal{C}^{j_n}(x_n)\bigg]e^{-s\frac{\sum_{n=1}^Nj_n\beta(x_n+1)}{2}},
\end{align}
where
$$
\eta=\binom{T}{j_1,...,j_N}=\frac{T!}{\prod_{n=1}^Nj_n!},
$$
and the set $\Phi$ is defined as
$$
\Phi=\bigg\{j_1,...,j_N|T=\sum_{n=1}^Nj_n, j_n=0,1,...,T\bigg\}.
$$
  By using the inverse Laplace transform of $[\mathcal{L}(s)]^T$ in \eqref{eqn:AD6}, the PDF of $Y$ can be obtained
\begin{align}\label{eqn:AD7}
&f_Y(y)=\mathcal{L}^{-}[[\mathcal{L}_{y_t}(s)]^T](y)\nonumber\\
&\approx c^T\sum_{\{j_1,...,j_N\}\in\Phi}\eta\bigg[\prod_{n=1}^N\mathcal{C}^{j_n}(x_n)\bigg]\mathcal{L}^{-1}\bigg(e^{-s\frac{\sum_{n=1}^Nj_n\beta(x_n+1)}{2}}\bigg).
\end{align}

Therefore, the outage probability of user 1 can be approximated as
\begin{align}\label{eqn:AD8}
P_{1,T}^{N}&=\int_0^{r_1}\mathcal{L}^{-}[[\mathcal{L}_{y_t}(s)]^T](y)dy\nonumber\\
&\approx c^T\sum_{\{j_1,...,j_N\}\in\Phi}\eta\bigg[\prod_{n=1}^N\mathcal{C}^{j_n}(x_n)\bigg]\nonumber\\
&\cdot\underbrace{\int_0^{r_1}\mathcal{L}^{-1}\bigg(e^{-s\frac{\sum_{n=1}^Nj_n\beta(x_n+1)}{2}}\bigg)(y)dy}_{Q_{r_1}}.
\end{align}
Using Gaver-Stehfest procedure \cite{Uni2006} and Gaussian-Chebyshev quadrature, we have
\begin{align}\label{Qfun}
   Q_{r_1}\approx \sum_{m=1}^N \frac{\pi\ln2\sqrt{1-x_m^2}}{Nr_1(1+x_{m})}\sum_{k=1}^L \hat{w}_ke^{-\frac{\sum_{n=1}^Nj_n\beta(x_n+1)k\ln2}{r_1(1+x_{m})}}.
\end{align}
Substituting \eqref{Qfun} into \eqref{eqn:AD8} completes
the proof.



\end{document}